\documentclass[prx,twocolumn,a4paper]{revtex4-2}

\usepackage{geometry}                		
\usepackage{graphicx}				
\usepackage{amsmath}
\usepackage{amssymb}
\usepackage{color}
\usepackage{hyperref}
\usepackage{ascii}
\usepackage{tikz}
\usetikzlibrary{quantikz2}
\usepackage{adjustbox}

\begin{document}
\title{Optimization Driven Quantum Circuit Reduction }

\author{Bodo Rosenhahn$^1$, Tobias J.\ Osborne$^2$, Christoph  Hirche$^1$}
\address{1) Institute for Information Processing (tnt/L3S), Leibniz Universit\"at Hannover, Germany}
\author{ }
\address{2) Institute for Theoretical Physics (ITP/L3S), Leibniz Universit\"at Hannover, Germany}


\date{\today}

\begin{abstract}
Implementing a quantum circuit on specific hardware with a reduced available gate set is often associated with a substantial increase in the length of the equivalent circuit. This process is also known as transpilation and due to decoherence, it is mandatory to keep quantum circuits as short as possible, without affecting functionality. 
In this work we propose three different transpilation approaches, based on a localized term-replacement scheme, to substantially reduce circuit lengths while preserving the unitary operation implemented by the circuit. The first variant is based on a stochastic search scheme, and the other variants are driven by a database retrieval scheme and a machine learning based decision support. We show that our proposed methods generate short quantum circuits for restricted gate sets, superior to the typical results obtained by using different qiskit optimization levels. Our method can be applied to different gate sets and scales well with an arbitrary number of qubits. 
\end{abstract}

\maketitle

\section{Introduction}
As the realisation of a fully fault-tolerant quantum computer capable of supporting error-free logical qubits and gates draws ever nearer, it is vital that methods to optimize the length of quantum circuits be investigated. The reasons here are twofold: (1) just below the fault tolerance threshold decoherence will still play a critical role and shorter quantum circuits will suffer less; and (2) just above threshold the implementation of a logical gate is likely to still be very expensive in terms of physical qubits and gates. Thus the development of schemes to discover optimal quantum circuits are a priority. The automated search for optimal quantum circuits to implement a target unitary is known as \emph{quantum architecture search} (QAS). The name is motivated by terminology arising in the machine learning community, where \emph{neural architecture search} \cite{Miikkulainen2020,xie2018snas} deals with algorithm selection and its hyper-parameter tuning. Common approaches are based on reinforcement learning, structural search or performance prediction, as presented in \cite{BakerGRN18,CaiCZYW18,negrinho2019towards}. 

QAS is often based on discrete and heuristic optimization strategies as the optimization criteria can be non-differentiable. Here, for example,
Gibbs sampling 
\cite{PhysRevResearchLi20}, evolutional approaches  \cite{9870269}, genetic algorithms \cite{RasconiOddi2019,PhysRevLett116.230504}, neural-network based predictors \cite{Zhang2021}, variants with noise-aware circuit learning \cite{PRXQuantum21}, and the optimization of approximate solutions \cite{PhysRevX10021067}  have been suggested.  Several works also demonstrated that  gradient-descent based optimization schemes \cite{quantum3020021,Zhang2022}, Monte Carlo Tree Search (MCTS) \cite{WangAQC23},  Monte Carlo Graph Search
\cite{RosOsb2023a,PhysRevA.110.022443}, ranking schemes \cite{He_Deng_Zheng_Li_Situ_2024}, and reinforcement learning \cite{PhysRevResearch.2.033446} are promising strategies.  Recent surveys on QAS are provided in  \cite{Zhu23ICACS,martyniuk2024quantumarchitecturesearchsurvey}.

The closely related task of translating, or \emph{transpiling}, a given quantum circuit into a sequence of gates realizable on a specific hardware architecture is known as \emph{Quantum Architecture Mapping} (QAM) \cite{Datta22, Stefano2024}. Realising QAM brings about four key challenges, 
(a) \emph{gate synthesis}, to convert gates or gate sequences into a set of gates supported
by the target hardware; (b) \emph{gate mapping}, to rearrange the qubits in the circuit to match the connectivity constraints of
the target hardware; (c) \emph{gate optimization}, to minimize the overall number of gates in the
circuit; and (d) \emph{the resource allocation}, to ensure that the circuit does not exceed the resources available on the
target hardware. Often, QAS, QAM, or transpilation, involves methods which have been motivated from electronic design automation (EDA) \cite{Top23}.  It should also be noted, that these terms are often used inconsistently in the literature.

\begin{figure}
\centering
  \includegraphics[width=0.5\textwidth]{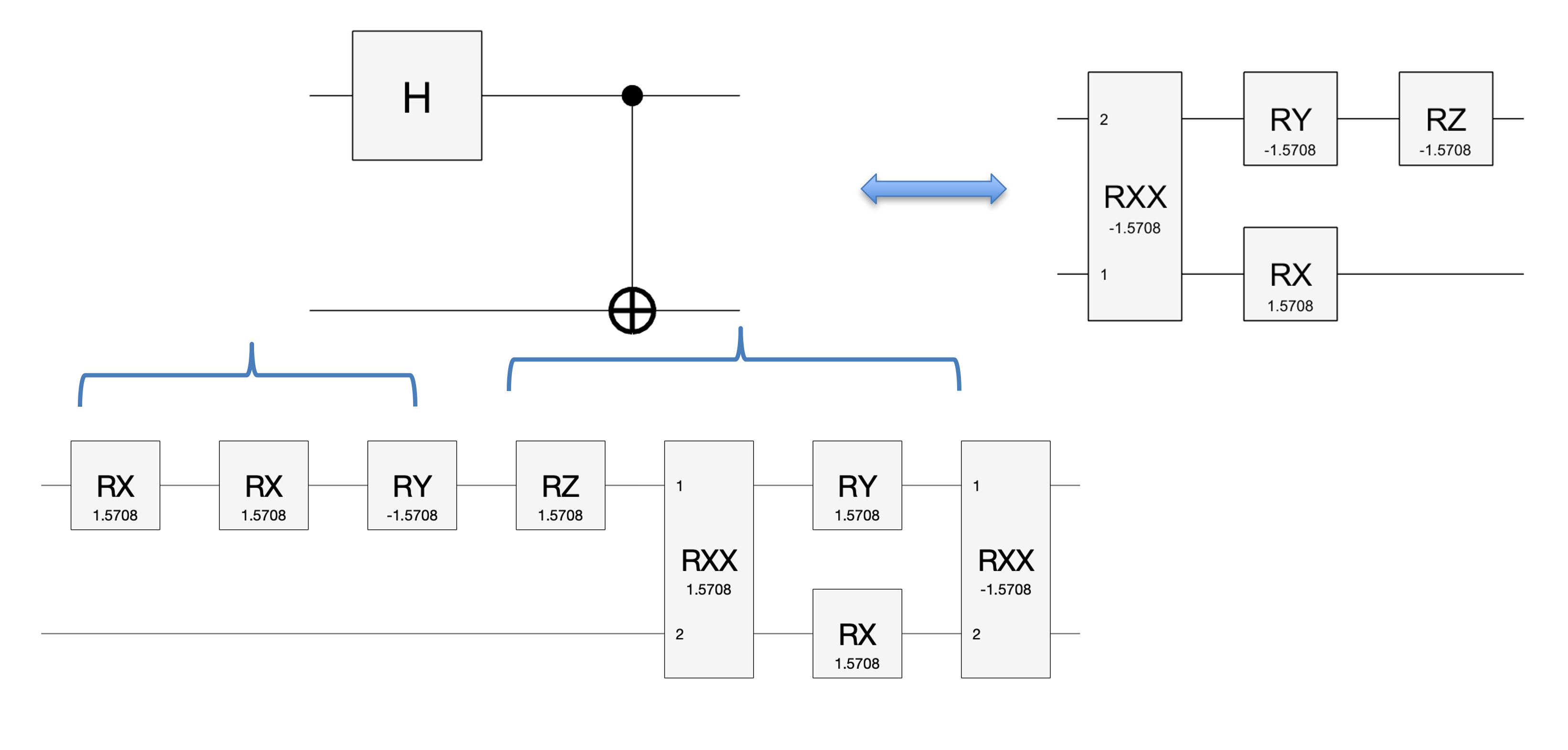}    
\caption{Motivation: The quantum circuit to produce a Bell state (H-Gate followed by a CNot Gate).  Left: Naive quantum architecture mapping to an ion trap quantum computer gate set, by mapping one operator after another. Right: Resource efficient (and equivalent) quantum circuit.
}
\label{fig:Mot}
\end{figure} 

There have been a variety of papers investigating QAM. In \cite{Datta22} the 2-dimensional square, heavy-hex and fully hexagonal qubit coupling lattices are considered as targets for transpiled quantum circuits. Implementing quantum algorithms on multi-core quantum computing architectures was discussed in \cite{Ovide23} and \cite{zulehner2018mapping}, which present a methodology for mapping quantum circuits to the IBM-QX architectures. Brandhofer et al.\ \cite{9643490} address the primary objective of adapting a quantum circuit to the topology of a provided quantum computer so that the qubit-qubit interaction requirements of each computation are satisfied. To adapt to the topology, several {\sc swap} gate insertion models are optimized using a Z3 SMT solver \cite{deMoura08}. The authors carried out numerical experiments on quantum circuits involving up to 15 qubits and a maximum depth of 76. In \cite{9138945} the architecture mapping for trapped ion (TI) qubits is addressed. Besides single traps, so-called Quantum Charge Coupled Device (QCCD) architectures allow the use of several traps in parallel. This poses optimization challenges for the shuttling, split/merge, and swap operations required to perform quantum computations. Most of these works rely on evolutionary optimization \cite{Dat22}, reinforcement learning \cite{Li24}, and similar techniques as they are used in quantum architecture search. The present work is mainly inspired by \cite{RosOsb2023a, 9643490}, but our focus is on circuit reduction and we therefore assume an existing mapping as an upper bound on the circuit length, and we propose the use of local term replacement operators to reduce the circuit length iteratively.  

A key challenge for quantum architecture mappings is that only a handful of gates are provided by the hardware. Thus the naive translation of individual blocks, followed by the assembly of the blocks, one after the other, can lead to vastly long quantum circuits exhibiting many redundancies and inefficiencies. Due to decoherence and/or the overhead of quantum error correction, it is vital to find an efficient and shorter equivalent quantum circuit to obtain reliable results. This is especially important for, e.g., quantum volume metrics \cite{Ryabov15}, a metric that measures the capabilities and error rates of a quantum computer. It is
typically used to evaluate and compare quantum computers across different platforms. Figure \ref{fig:Mot} illustrates the challenges arising in realising QAM in terms of a small example to prepare a Bell state  using elementary realizable quantum gates (left). An optimized circuit is depicted on the right. 

In the classical literature the analogous problem to QAM is related to \emph{logic minimization}, which addresses the task of replacing a group of (logic) gates with another (smaller group) that will perform the same task faster or by using less space on a chip.  A standard motivation is that the equivalent optimized circuit is cheaper, more compact on a chip, more energy efficient, has reduced latency, and it minimizes risks of cross-talk. Karnaugh maps \cite{Marcus1967} are a typical device to carry out such logic minimization. For 
Karnaugh maps, the idea is to order the constraints via a so-called Gray code. A Gray code ensures that only one variable changes between each pair of adjacent cells. The minimal terms for the final expression are found by encircling groups of 1s in the Karnaugh map which provide the best opportunities to simplify the expression. Unfortunately, this approach can not be directly applied to quantum circuits.

To address the challenges of realizing large-scale automated QAM we propose, in this paper, the use of stochastic, database and machine learning (ML) supported reduction methods to identify redundancies and algebraic manipulations to efficiently reduce the circuit length.  Our core contributions are three different methods to perform quantum circuit reduction, and the evaluation of their impact on quantum hardware:
\begin{enumerate}
\item Method 1 (V1-RS): We propose a simple sub-block selection and random sampling (RS) based term replacement algorithm to identify reducible circuit blocks in a quantum circuit.
\item Method 2 (V2-DR): We study the precomputation of a compute graph of predefined depth to derive a database of optimal (yet still reasonably small) quantum maps. Then a random selected sub-block and database retrieval (DR) scheme is used to efficiently reduce these sampled circuit blocks.
\item Method 3 (V3-RF) is built on V2, where we let a classifier make a decision if a sampled sub-circuit block is likely reducible. In our experiments, the decision is made by a random forest (RF), which is very fast to compute at inference time. Only after the classifier decision is the database retrieval scheme called. This prevents unnecessary database look-ups which can be time consuming.
\item An evaluation on real quantum hardware reveals that the reduced quantum circuits have a significant impact on the results. Our approach is also superior to different qiskit optimization levels.
\item The source code for the graph generation and quantum circuit length optimization will be made publicly available.
\end{enumerate} 

\section{Preliminaries}
\subsection{Quantum Gates and Circuits}
For the sake of simplicity we assume that a target quantum computer is comprised of $N$ \emph{logical} qubits, forming a quantum register \cite{10.5555/1206629}. 
Thus the Hilbert space appropriate for our system is given as $\mathcal{H}\equiv (\mathbb{C}^2)^{\otimes N} \cong \mathbb{C}^{2^N}$. The axioms of quantum mechanics posit that quantum logic gates are unitary matrices. Thus, a gate acting on $N$ qubits is represented by a $2^{N}\times 2^{N}$ unitary matrix. A quantum gate sequence is then simply a set of such gates which are, in turn, evaluated via a series of matrix multiplications: A quantum circuit of length $L$ is thus described by an ordered tuple $(O(1), O(2), \ldots, O(L))$ of quantum gates and the corresponding unitary operation $U$ is given by the product 
\begin{equation}
  U = O(L)O(L-1)\cdots O(1).  
\end{equation}  
Standard elementary quantum gates often involve the Pauli-($X$, $Y$, $Z$) operations, as well as \mbox{Hadamard-}, $\textsc{cnot}$-, $\textsc{swap}$-, phase-shift-, and $\textsc{toffoli}$-gates, all of which are expressible as standardised unitary matrices~\cite{nielsen2010quantum}. 
The action of a quantum gate is extended to act on a register of any size by making use of a tensor product with the identity operator.

We interpret each quantum circuit as a token chain, e.g.\
$  U = O(L)O(L-1)\cdots O(1) $ = {\tt O(L) O(L-1) \dots O(1) },
where each token represents an operator of the quantum circuit. As some operators are commutative it is possible to swap specific tokens without changing the resultant unitary implemented by the quantum circuit; we will perform term replacement on this token chain to reduce the length of an existing quantum circuit without changing its functionality.

\subsection{Compute Graphs}
\label{SecCompGraph}
If we are provided a limited set of available gates then it is possible to explore all possible implementable quantum circuits by building the so-called \emph{compute graph} up to a predefined depth. These are combinatorial objects with nodes given by quantum circuits and edges labelled by elementary gates.

\begin{figure}
\centering
\includegraphics[width=0.5\textwidth]{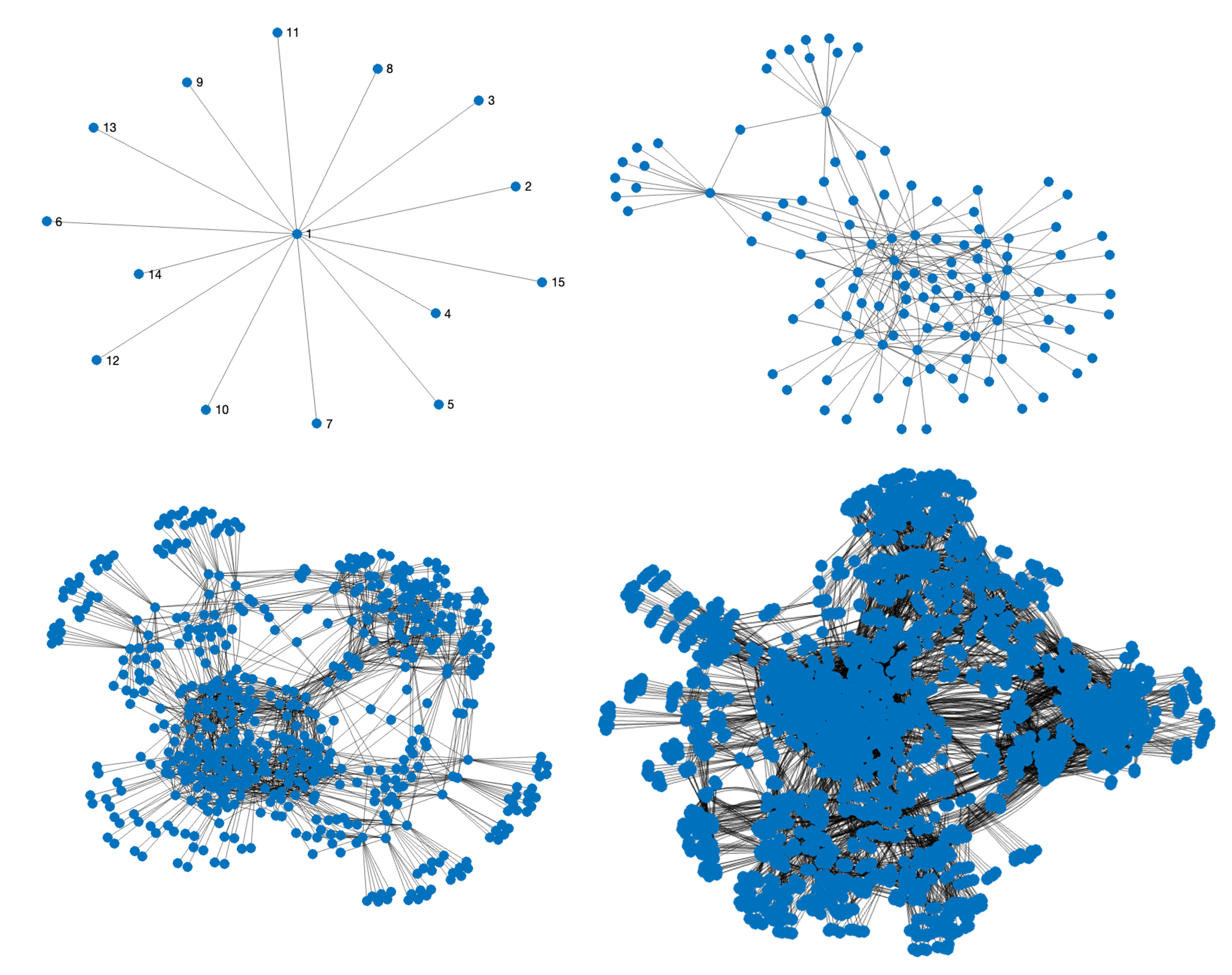}    
\caption{Compute Graphs of depth 1 to 4 (from the upper left to the lower right) for a provided set of available quantum operators. 
}
\label{fig:FCG}
\end{figure}

\begin{table}
\centerline{
\begin{tabular}{|c|c|c|c|c|}
\hline
Qubits & $\#$Operators & Depth & Nodes & Edges\\
\hline
2 & 14 & 1 & 15 & 14 \\
2 & 14 & 2 & 114 & 210 \\
2 & 14 & 3 & 584 & 1596 \\
2 & 14 & 4 & 2024 & 8176 \\
2 & 14 & 5 & 4512 & 28336 \\
2 & 14 & 6 & 7420 & 63168 \\
\hline
3 & 24 & 1 & 25 & 24 \\
3 & 24 & 2 & 337 & 600 \\
3 & 24 & 3 & 3215 & 8088 \\
3 & 24 & 4 & 23622 & 77160 \\
3 & 24 & 5 & 137572 & 566928 \\
\hline
 \end{tabular}}
 \caption{Full Compute Graphs for different depths and amount of operators }
 \label{tab:QGD}
 \end{table} 
In our preliminary work
\cite{RosOsb2023a} we made use of such a compute graph to perform quantum architecture optimization via Monte Carlo Graph Search (MCGS). The method starts
 with the identity operator $ \mathbb{I}$ as root node and iteratively builds up quantum circuits by selecting gates from 
a predefined set $\mathcal{OP}=\{O_1, O_2, \ldots\}$ of elementary quantum gates.
The compute graph is then a graphical model with nodes containing unitary matrices and edges encoding an elementary unitary operator $O_i \in \mathcal{OP}$. The graph is initialized with the identity matrix $ \mathbb{I}$ as root node. An operator $O_i$ is selected and applied to the root node. This yields  a new node by multiplying the selected operator
with the unitary matrix of the parent node. If the resulting unitary already exists as node in the graph, a direct edge from the parent to the already existing node can be added. Otherwise, a new node is generated and connected with the parent node. Please note that we compare unitaries numerically with a tolerance of $10^{-5}$ and we compensated for a global phase.
While growing the graph, the resulting unitary matrices are provided as graph nodes and the underlying quantum circuit can be computed by finding the shortest path from the root node to the target unitary and by collecting the operators along the edges of the path. 
Figure \ref{fig:FCG} shows the emerging graph with increasing depth and we refer to \cite{RosOsb2023a} for further details.

Thus, each node is identified with a possible quantum circuit. It is notable that this graph contains cycles since identical quantum circuits have multiple representations in terms of different gates and gate orders. As the graph will increase exponentially, it is only feasible to build up a full graph for reasonably short circuits. For MCGS it is therefore important to restrict the growth direction in an unbalanced fashion to explore graph structures which are more likely to be useful for solving a certain task. Poisson sampling can exploited as the underlying sampling process to select a vertex to further develop the current compute (sub)graph. It is the basic paradigm of
Monte Carlo Search 
\cite{doi:10.1080/01621459.1949.10483310} and adapted Gibbs sampling \cite{george1993variable}  to iteratively grow a graph.
In contrast to the former work, we fully grow the graph to a certain depth and collect for all nodes the shortest path to the root as the most efficient quantum circuit implementation of the node unitary. The table \ref{tab:QGD} summarizes how the graph increases with increasing depth. Note that several cycles in the developed graph exist, and for a pool of e.g.\ $n$ operators, a depth of $m$ leads to far fewer nodes than $n^m$. This is the main reason for the efficiency of MCGS compared to classical MCTS models.
Figure \ref{fig:FCG} shows the compute graphs for $14$ predefined operators along the depth 1, 2, 3 and 4. Note, that this graph can be pre-computed and resulting unitary matrices (on the nodes), as well as their perfect factorization, can be stored in a database.
This will be later used (V2) to obtain a more efficient term replacement scheme compared with a stochastic random search.
\subsection{Random Forests}
\label{Sec:RF}
As our last variant V3 is be based on a fast machine-learning based decision. To elucidate this scheme we first summarize in this section the principle of a decision tree, and based on this, a random forest. 
A decision tree is a hierarchical model performing splits based on local decisions. Given a training data set, the goal is to find a splitting criterion that maximizes the information gain by measuring the entropy of the parent (before the decision) and comparing it to the entropies of the children (after the decision).
More formally, the goal of a split node is to maximize the information gain of the decision. We exploit the Shannon entropy, defined as 
\begin{equation}
   \mathrm {H} (T)=\operatorname {I} _{E}\left(p_{1},p_{2},\ldots ,p_{J}\right)=-\sum _{i=1}^{J}p_{i}\log _{2}p_{i},
   \end{equation}
where $ p_{1},p_{2},\ldots$  are probabilities that add up to 1 that represent the percentage of each class present in the data set. Then the information gain of a splitting can be expressed as the sum of entropies of the children nodes subtracted from the entropy of the parent node.
Thus, the information gain at a node $T$ using attribute $a$ can be measured as
\begin{eqnarray*}
  \underbrace {IG(T,a)}_{\text{info gain}}&=&\underbrace {\mathrm {H} (T)}_{\text{Ent (parent)}}-\underbrace {\mathrm {H} (T\mid a)}_{\text{Ent (children)}}\\ 
  =-\sum _{i=1}^{J}p_{i}\log _{2}p_{i}&+&\sum _{i=1}^{J}\Pr(i\mid a)\log _{2}\Pr(i\mid a) 
\end{eqnarray*}
where $H(T|a)$  is the conditional entropy of $T$ given the value of attribute  $a$. 
The information gain is used to identify the perfect splitting nodes to build up a decision tree. We refer to \cite{Quinlan1986InductionOD} for further details.

\begin{figure*}
\centering
  \includegraphics[width=0.95\textwidth]{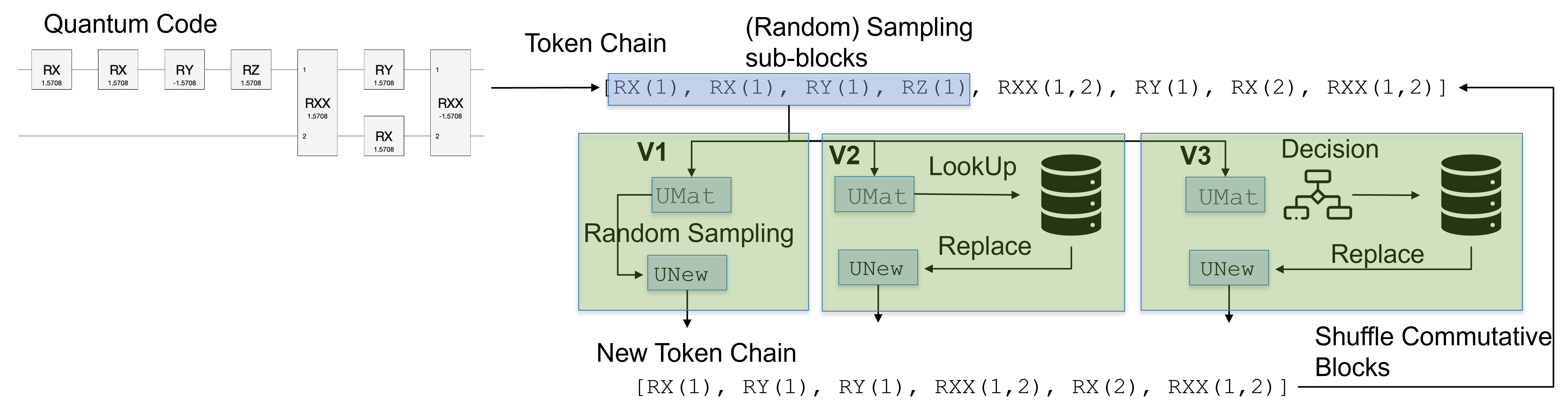}    
\caption{Variants of the quantum circuit optimization process: V1 is based on simple random sampling (RS), V2 is based on a database retrieval (DR) by using compute graphs and V3 is based on a decision scheme whether it is worth to perform the database lookup (RF).
}
\label{fig:Var3}
\end{figure*} 

Based on the decision scheme a random forest can be generated as an ensemble of several (different) decision trees
\cite{Breiman01}.
To ensure different decision trees, bootstrap aggregation and bagging is used, which means that for each decision tree only a random subset of the training data and available features is used and the final decision is based on the average (regression tree) or majority vote (classification tree)
\cite{BreiFrieStonOlsh84}. Please note, that the training of a random forest is very fast, e.g. the most complex models we used in the below experiments were trained in less than a second on a simple notebook.
The inference is much faster (below 0.0001 seconds).
\section{Proposed Methods}

This section summarizes the three proposed methods for iterative quantum circuit reduction. Figure \ref{fig:Var3} provides a general overview of the method and indicates the three proposed variants. 

\subsection{V1 : Random Search}
As mentioned above, we interpret each quantum circuit as a token chain, e.g.
$  U = O(L)O(L-1)\cdots O(1) $ = {\tt O(L) O(L-1) \dots O(1) }
where each token represents an operator of the quantum circuit.
In theoretical computer science, 
 a formal language consists of words whose letters are taken from an alphabet and are well-formed according to a specific set of rules called a formal grammar \cite{Hopcroft79}. A formal grammar mainly consists of a set of production rules and rewriting rules for transforming strings. Each rule specifies a replacement of a particular string  with another. Several books cover this fundamental topic, e.g.
  \cite{Leuuwen91}.

\begin{figure}
\centering
\includegraphics[width=0.5\textwidth]{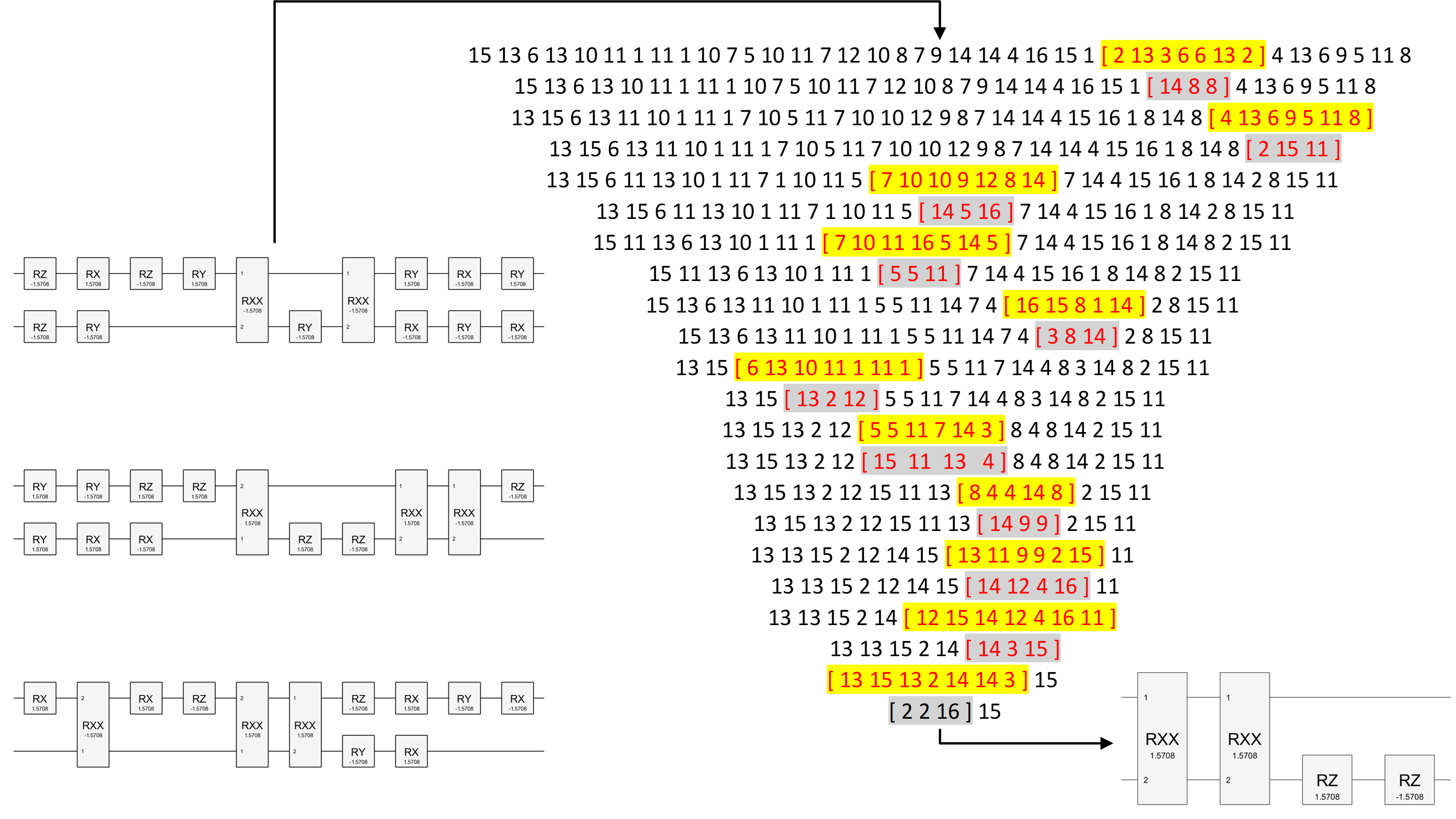}  
\caption{Example optimization steps of an input circuit length with 40 operators and intermediate reduction steps during the iterations until a substantially reduced circuit  results.
}
\label{fig:InOutToken2}
\end{figure} 

The random search algorithm works as follows: Given a token chain $ U= O(L) O(L-1) \dots O(1) $, first, a random connected subset, e.g. at position $m$ of length $n$ is selected, 
\begin{eqnarray*}
U &=& O(L) \dots \underbrace {(O(m+n-1) \dots O(m))}_{\text{length n}} \dots  O(1)\\
 &=& O(L) \dots (U_s) \dots  O(1).
\end{eqnarray*}
The subset $ O(m+n-1) \dots O(m) $ generates a unitary  $U_s$.
Then random sets of $p$ tokens $O^r(i)$ with $p<n$ are sampled and the obtained unitary is compared with the unitary of the selected subset, $ U_p = O^r(p) \dots O^r(1)$.
If $U_p=U_s$  (up to a global phase and within a tolerance of $10^{-5}$), the token chain can be replaced by a shorter one, 
\begin{eqnarray*}
U &=& O(L) \dots \underbrace {(O(m+n-1) \dots O(m))}_{\text{length n}} \dots  O(1) \\
&=&  O(L) \dots (U_p) \dots  O(1) \\
&=& O(L) \dots \underbrace {(O^r(p) \dots O^r(1))}_{\text{length p}} \dots  O(1)
\end{eqnarray*}
Finally, commutative blocks are randomly exchanged and the process is iterated.
Figure \ref{fig:InOutToken2} shows the process for an input circuit of length 40, its translation to a token chain (as numbers), and the individual term replacements which were detected, until a reduced gate set is reached.

\subsection{V2 : Compute graph database retrieval}

\begin{figure}
\centering
\includegraphics[width=0.5\textwidth]{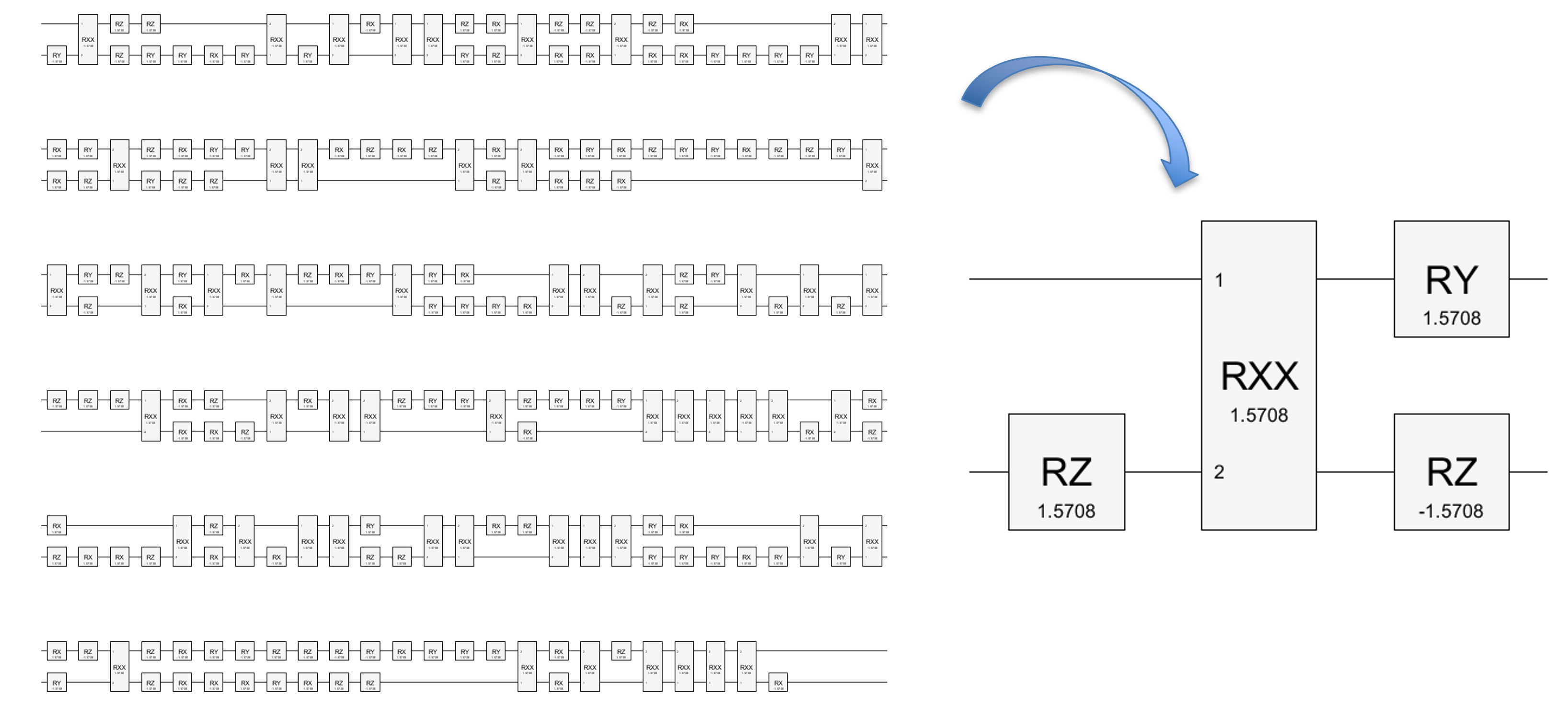}  
\caption{Example optimization of an input circuit length with 200 operators and a substantially reduced circuit after the term replacements.
}
\label{fig:InOut2}
\end{figure}

Variant 2 is an extension of the random search strategy in V1. A major disadvantage of V1 lies in the two nested stochastic processes required to perform the random selection of the token chain to be optimized and the random sampling to find more efficient token chains. This can lead to a large computation time. One observation we made is that many blocks (e.g. of length five to eight) are often reduced to lengths between 2 and 4. This, in turn, is a circuit length which is feasible to fully compute within a compute graph, as presented in section \ref{SecCompGraph}. Thus, the idea is to collect all nodes with its corresponding unitary of a compute graph and to compute its shortest path as its most efficient factorization and to store them in a database, e.g.
\begin{eqnarray*}
U_1 & \rightarrow & O^1(n_1) \ldots O^1(1) \\
U_2 & \rightarrow & O^2(n_2) \ldots O^2(1) \\
&\vdots& \\
U_m & \rightarrow & O^m(n_m) \ldots O^m(1) \\
\end{eqnarray*}
Table \ref{tab:QGD} shows how the size  of this database (the amount of Nodes) increases, even though when only a small number of operators and a small depth is chosen.  Still, e.g., a lookup and comparison with 20K entries when there are 24 operators and a full depth of 4 is used, is manageable. More importantly, after the lookup, there is a guarantee whether the selected token chain can be reduced or not, which avoids unnecessary redundant compute steps.
As indicated in \ref{fig:Var3}, the proposed V2 replaces the random sampling of subblocks with a database lookup. The experiments indicate that this can significantly reduce the computation time, see e.g.\ Figure \ref{fig:Conv1}. Figure \ref{fig:InOut2} shows an example of an optimization with an input circuit length of 200 tokens which has been significantly reduced using this iterative optimization scheme.

\subsection{V3 : Random forest supported database lookup}
The optimization version V2 relies on a database lookup. Unfortunately the database can become very large with an
increasing number of operators and increasing depth (see Table \ref{tab:QGD}). Given a randomly selected token chain, we now propose to train a small classifier (here, we make use of a random forest) to decide if it is worthwhile performing a database lookup. This classifier will be trained on the nodes of a compute graph to learn irreducible circuit blocks and successfully reduced circuit blocks from V2 as reducible examples. E.g. on our current implementation, the database lookup takes around 0.004 seconds, whereas a random forest can be evaluated in 0.0001 seconds. Thus, a prefiltering based on the random forest can prevent unnecessary computations. The optimization of a random forest is summarized in Section \ref{Sec:RF} and can be computed within seconds on a standard laptop. The random forest is optimized beforehand and then used as existing method in our pipeline. Our software examples also provide the code for training a random forest and is therefore fully reproducible. As indicated in Figure~\ref{fig:Var3}, the proposed method V3 adds in front of the database lookup a small random forest. 

\section{Experiments}
In our experiments we demonstrate the efficiency of our proposed approaches to reduce a quantum circuit. The Figures \ref{fig:Mot} and \ref{fig:InOut2} show two practical results for automated quantum circuit reduction as a first proof of concept.
We focus in our experiments on the few qubit regime, as here the correlation between the operators is more complex and intertwined than for a larger number of qubits. A large number of qubits leads to parts of the circuit becoming independent with increased number of qubits. Further several parts of the circuit are more likely to be irreducible. Therefore, the reduction of three or four qubit circuits can be more challenging and difficult than for a 20 qubit code block as more algebraic manipulations are required to efficiently reduce the blocks. We address the aspect of scaling to a larger qubit number in section \ref{Sec:Impact}, separately.
 
\begin{figure}
\centering
  \includegraphics[width=0.5\textwidth]{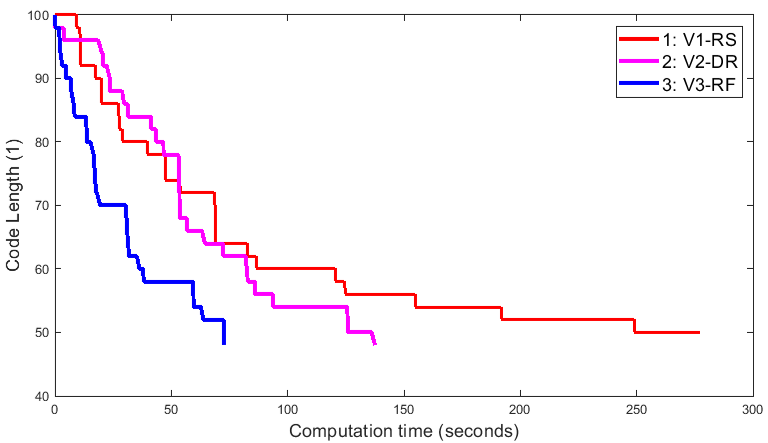}    
\caption{Quantum circuit reduction. The x-axis shows the computation time and the y-axis the circuit length which has to be reduced. The random forest supported reduction scheme performs fastest.
}
\label{fig:Conv1}
\end{figure} 

Figure \ref{fig:Conv1} shows the performance of the three proposed algorithms on an example quantum circuit. In this case, the task is to reduce the circuit length of 100 to around 50. The x-axis shows the computation time in seconds and the y-axis the circuit length which is iteratively reduced during the iterations. As can be seen, all three proposed algorithms successfully converge, but the variants V2 and V3 are more efficient, with V3 being the fastest version. 
As mentioned before, for most of our experiments, we restrict the amount of qubits to two and three. This implies, that all gates are highly connected and thus the optimization and algebraic manipulations to reduce the code are correspondingly harder. In section \ref{Sec:Impact} we present an approach to upscale our method to an arbitrary number of qubits, while keeping the size of the compute graph constant.

\begin{figure}
\centering
  \includegraphics[width=0.5\textwidth]{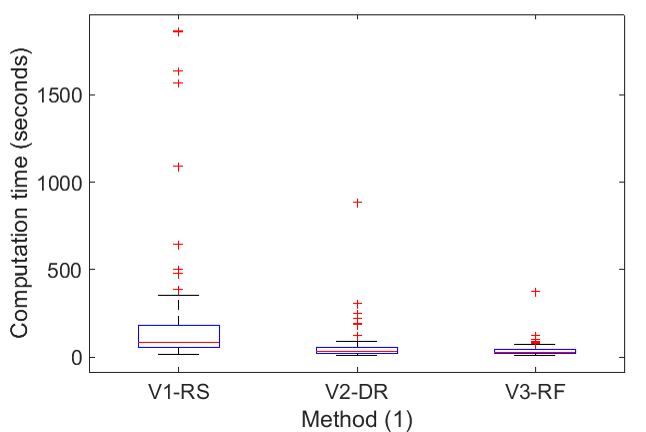}    
\caption{Statistical summary of the three proposed methods. 100 different randomly sampled quantum circuits are reduced to a predefined fixed length.
The y-axis shows a box-plot. In each box, the central mark indicates the median, each box indicates the 25th and 75th percentiles. The whiskers extend to the most extreme data points not considered outliers, and the outliers are plotted individually using the + marker symbol.
}
\label{fig:Stat1}
\end{figure} 

\begin{table}
\centerline{
\begin{tabular}{|c|c|c|c|}
\hline
(100 runs) & V1-RS & V2-DR & V3-RF\\
\hline
Mean & 199 (sec) & 55 (sec) & 38 (sec) \\
Stddev & 351.5 & 96.3 & 39.8 \\
\hline
 \end{tabular}}
 \caption{Summarizing statistics on the computation time of the three proposed methods for 100 randomly generated quantum circuits, as in Figure \ref{fig:Stat1}.  }
 \label{tab:Stat1}
 \end{table} 

As all three algorithms are based on a stochastic process (of random sub block selection), we further evaluated the performance in Figure \ref{fig:Stat1} and table \ref{tab:Stat1}. For this experiment we randomly generated 100 different starting circuits of length 100 and reduced the circuit to a target size. Then we measure the mean and variance of all three variants over this large amount of repetitive experiments. As can be seen in table \ref{tab:Stat1}, the performance of the three proposed variants can be summarized as V3 performing as best algorithm (as the fastest one with the least variance in results) and V1 performs worst. 
A statistical summary is also provided in Figure \ref{fig:Stat1} by using a box-plot for each method. The box-plot shows the median, the 25th and 75th percentiles as well as the most extreme data points (whiskers) and outliers.
It should be noted, that we excluded the computation time to compute the graph and to train the small classifier as both parts can be done offline. Whereas V1-RS requires in average 199 seconds, already the database lookup reduces the compute time to nearly a quarter (55 seconds). The variant V3 making use of the random forest brings the average computation time down to 38 seconds. Also the standard deviation is lower for V3, which indicates a reliable convergence behavior over time.
\subsection{Impact and Scalability}
\label{Sec:Impact}
\begin{figure}[ht]
\centering
\includegraphics[width=0.45\textwidth]{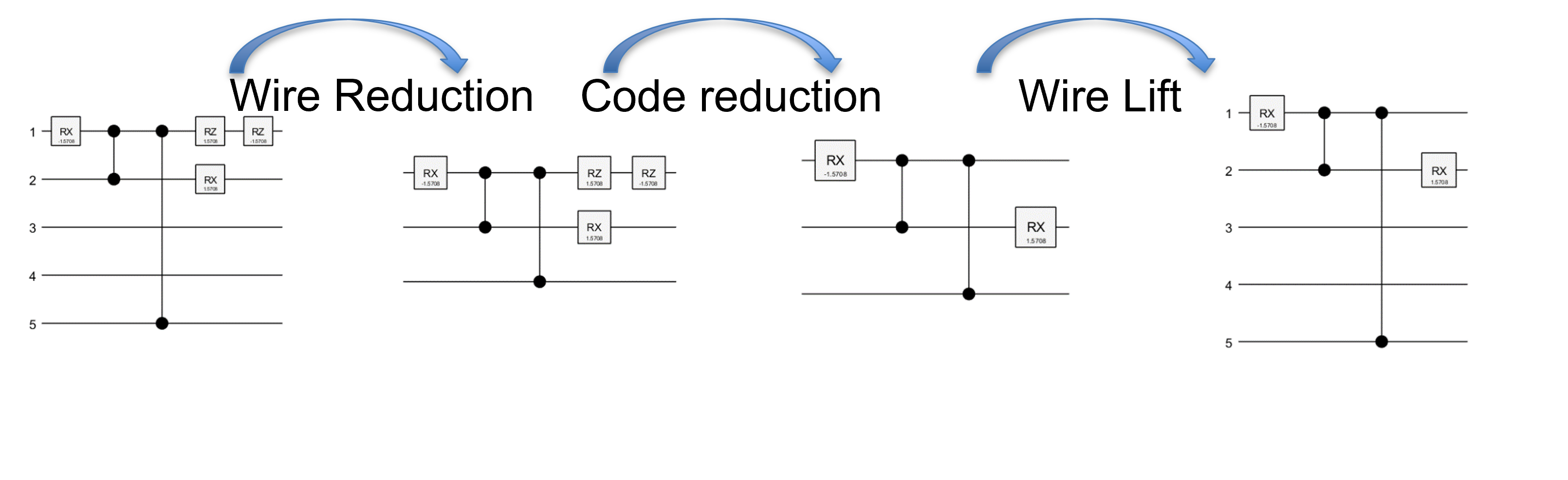}    
\caption{Principle for scaling up the number of qubits: Since only short block segments are selected for reduction, non-used qubits an be removed, the optimization can be done on the lower dimensional space and the lifted up again to the original code size.
}
\label{fig:Scalability}
\end{figure}

The complexity of a compute graph increases exponentially with the number of qubits, the available gates, possible gate combinations. Thus the presented approach is not suitable for a naive upscaling. To still achieve circuit optimization for a larger number of qubits (e.g.\ beyond four or five), the general idea and observation is that, if only short circuit blocks are selected (e.g. of length between three to seven), usually many wires are not connected to these gates and they only cause an increase of the underlined dimensions. Thus, we reduce the non-needed wires and map the selected gate sequence to a subspace only containing required qubits.
 After the analysis in the smaller qubit space and potential block
 reduction, the resulting circuit is mapped back to the original size. Figure \ref{fig:Scalability} visualizes the general idea. This approach allows us to deal with arbitrary sizes, in the latter experiments we show results with up to 15 qubits. Please also note that, in general, non-connected circuit blocks (e.g. with differently involved qubits) can be optimized in a parallel framework, which can speed up the optimization for complex architectures. We also observed that the compressing rate for larger qubit numbers decreases as the likelihood for non-reducible code blocks increases. Thus, our algorithm is best suited for very long circuits with a smaller amount of qubits as then more algebraic manipulations are necessary to reduce the circuit length. These more complex manipulations are not well covered by the lvl 1-3 optimizers of qiskit which explains our superior performance in the latter experiments.

Our approach is applicable to arbitrary (discrete) gate sets. To illustrate this we now present optimized circuit maps for an ion-trap architecture consisting of $RX, RY$, $RZ$ gates and $RXX$ gates as well as  for NISQ-architectures as provided by IBM, consisting of 
$RX, RY$ and $CX$ gates.
The proposed method is compared with a qiskit transpiler on optimization levels one to three. 
The description in qiskit only states vague information, e.g. for its highest level: \textit{Level 3 pass manager: significant optimization by noise adaptive qubit mapping and
    gate cancellation using commutativity rules and unitary synthesis} \footnote{ \url{https://github.com/Qiskit/qiskit/blob/main/qiskit/transpiler/preset_passmanagers/level3.py} (accessed 10/2024)}. 
    Tables \ref{tab:IOComp}, 
    \ref{tab:nisqComp} and  \ref{tab:nisqComp2}
    summarize results and 
show that our proposed method is  superior compared to all optimization levels, but it should be noted that the qiskit transpilation works very efficiently and  takes under a second, whereas our optimizer is takes much longer to terminate (see Table \ref{tab:Stat1}). 
\begin{table}
\begin{tabular}{|c|c|c|c|c|}
\hline
Method & $\#$ rx $\downarrow$ &   $\#$ ry $\downarrow$& $\#$ rz $\downarrow$ & $\#$ rxx $\downarrow$ \\
\hline
original & 86 & 88 & 86 & 40 \\
qiskit (lvl 1) & 27 & 30 & 55 & 40 \\
qiskit (lvl 2) & 22 & 41  & 83 & 35  \\
qiskit (lvl 3) & 21 & 40 & 81 & 34 \\
\hline
{\bf ours} & {\bf 5 } & {\bf 12} & {\bf 12} & {\bf 16} \\
\hline
\end{tabular}
\caption{Quantum circuit optimization example for an ion-trap architecture and comparison of our method to the  qiskit optimizer on levels 1-3 (3 qubits).}
\label{tab:IOComp}
\end{table}

\begin{table}
\begin{tabular}{|c|c|c|c|}
\hline
Method & $\#$ rx $\downarrow$ &   $\#$ rz $\downarrow$ & $\#$ cz $\downarrow$\\
\hline
original & 115 & 118 &  67 \\
qiskit (lvl 1) & 56 & 65 & 57 \\
qiskit (lvl 2) & 62 & 49   & 42 \\
qiskit (lvl 3) & 62 & 49& 42 \\
\hline
{\bf ours} & {\bf 32 } & {\bf 14} & {\bf 26} \\
\hline
\end{tabular}
\caption{Quantum circuit optimization example for a nisq architecture (IBM) and comparison of our method to the  qiskit optimizer on levels 1-3. 
(3 qubits). }
\label{tab:nisqComp}
\end{table}

\begin{table}
\begin{tabular}{|c|c|c|c|}
\hline
Method & $\#$ rx $\downarrow$ &   $\#$ rz $\downarrow$ & $\#$ cz $\downarrow$\\
\hline
original & 96 & 91 &  313 \\
qiskit (lvl 1) & 74 & 74 & 285 \\
qiskit (lvl 2) & 74 & 36   & 203 \\
qiskit (lvl 3) & 74 & 36& 203 \\
\hline
{\bf ours} & {\bf 72 } & {\bf 27} & {\bf 194} \\
\hline
\end{tabular}
\caption{Quantum circuit optimization example for a nisq architecture (IBM) and comparison of our method to the  qiskit optimizer on levels 1-3. 
(15 qubits). }
\label{tab:nisqComp2}
\end{table}

\subsection{Quantum Hardware Experiments}

\begin{figure}[ht]
\centering
\includegraphics[width=0.45\textwidth]{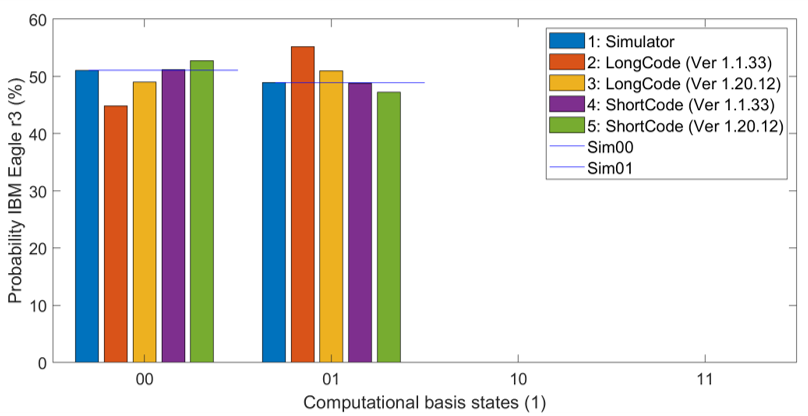}    
\caption{Measurement outcomes on two quantum chips (IBM) compared to the simulation result (most left).
We provide two equivalent quantum circuits, one is named \textit{long} for the sequence length of 40 and the other (equivalent on) is named \textit{short} with a sequence length of 8. Both circuits are equivalent and
the deviation from the simulation outcome (in blue) shows the importance to compile short circuits as decoherence over longer gate sequences is increasing significantly. 
}
\label{fig:QChipCompare}
\end{figure}

In the next experiment we generated a random two qubit quantum circuit of length 40 (long) and reduced it to a quantum length of 8 (short), keeping the implemented unitary intact. Thus both quantum circuits are equivalent, but we expect a higher noise ratio for the more inefficient quantum circuit.
Both quantum circuits have been transpiled in qiskit \cite{qiskit2024,QiskitTextbook:2020} and evaluated in the IBM quantum platform \footnote{ \url{https://quantum.ibm.com} (accessed 08/2024)}. Two different quantum chips have been used, both based on the IBM Eagle r3 architecture. This chip has 127 qubits, 5K CLOPS and  
they are named IBM:Brisbane and IBM:Kyiv on the platform. Whereas the IBM:Brisbane runs on version 1.1.33 with an EPLG (error per layered gate for a 100 qubit chain) of 4.17$\%$, the IBM:Kyiv processor runs on version 1.20.12 and has an EPLG of only  1.7$\%$ (according to the online documentation \footnote{ \url{https://quantum.ibm.com/services/resources?type=Eagle} (accessed 08/2024)}).
Figure \ref{fig:QChipCompare} summarizes the obtained measurements.
On the IBM:Brisbane, the long circuit has been transpiled to a circuit length of 156, whereas the short circuit has been transpiled  to just 34 circuit blocks. On the IBM:Kyiv it is slightly different as
the long circuit has been transpiled to a circuit length of 148, whereas the short circuit results in 32 circuit blocks.
 Figure \ref{fig:QChipCompare} shows the simulation (in blue) as left most bar and the different circuits on the two used processors in the other bars. 
The measurements in Figure \ref{fig:QChipCompare} show that the more recent version 1.20.12 provides a more stable outcome, compared to version 1.1.33, which is in accordance to the provided EPLG scores. Please note while this experiment is based on a 2-qubit circuit and a length of 40 gates which is nearly trivial, decoherence is still apparent and efficient transpiled quantum circuits are mandatory, now and in the future.

To summarize the experiments, our reduced quantum circuits are transpiled to significantly more efficient hardware implementable circuits and they lead to better results which are closer to the expectation provided by the simulator.

\section{Summary and Discussion}
Naively mapping a quantum circuit to an existing hardware can lead to a long quantum circuit with unnecessary redundancies.  Due to decoherence in quantum computers, it is mandatory to ensure we find equivalent, shorter and more efficient, circuits.
In this work we presented three different variants, based on a local term replacement scheme, to substantially reduce circuit length while maintaining the unitary implemented by the circuit. The first variant is based on a stochastic search scheme, the other variants are driven by a database retrieval scheme and a machine learning based decision support. We show that quantum circuit length can be efficiently reduced and 
different modifications can be done 
to significantly boost the compute time for optimization of the circuit length. 
It should be noted that our method is significantly slower than the qiskit transpilers and our approach is only useful when a highly efficient circuit block is required and reused several times in a larger context. It is also possible to perform simple reductions with the available transpilers and then to refine them further with our method. Thus, they are not exclusive to each other.  We also performed experiments using the ZX-calculus \cite{vandewetering2020zx}, but as discussed in other works, such as
    \cite{riu2024RLZX}, the outcome can be less efficient as the original input circuit. As we observed the same using ZX-calculus for our gate sets, there is need for further investigations and optimization, e.g. based on reinforcement learning. This will be part of future research. 
    We similarly experimented with the recently published IBM-AI model \cite{Kremer24IBM} for transpilation. It turned out 
    that the transpiler does not work with an arbitrary set of gates and it returned the level-3 transpiled code. We will address this as part of future research, as well.
In future works, we will also integrate costs for decoherence time of single operators and operator chains.  E.g. a token chain with five operators and two cx-gates might have a higher decoherence than a seven operator block with only one cx gate. With such cost measures we will replace our optimization criteria (which is just the code length at the moment) with alternatives in the future. 

\bigskip
\subsection*{Acknowledgments}
This work was supported, in part, by the Federal Ministry of Education and Research (BMBF), Germany under the AI service center KISSKI (grant no. 01IS22093C),
by the Quantum Valley Lower Saxony, by Germany's Excellence Strategy EXC-2122 PhoenixD, and the ERC and the DFG via the project ResourceQ. The authors would also like to thank the Qudora team for their valuable feedback. We also thank Ritajit Majumdar for the fruitful discussions and insights on the IBM transpiler toolset.
We acknowledge the use of IBM Quantum services for this work. The views expressed are those of the authors, and do not reflect the official policy or position of IBM or the IBM Quantum team.

\bibliographystyle{plain}
\bibliography{refs}

\begin{thebibliography}{10}

\bibitem{QiskitTextbook:2020}
Abraham Asfaw, Luciano Bello, Yael Ben-Haim, Sergey Bravyi, Lauren Capelluto,
  Almudena~Carrera Vazquez, Jack Ceroni, Frank Harkins, Jay Gambetta, Shelly
  Garion, Leron Gil, Salvador De La~Puente Gonzalez, David McKay, Zlatko Minev,
  Paul Nation, Anna Phan, Arthur Rattew, Joachim Schaefer, Javad Shabani, John
  Smolin, Kristan Temme, Madeleine Tod, and James Wootton.
\newblock Learn quantum computation using qiskit, 2020.

\bibitem{BakerGRN18}
B.~Baker, O.~Gupta, R.~Raskar, and N.~Naik.
\newblock Accelerating neural architecture search using performance prediction.
\newblock In {\em 6th International Conference on Learning Representations,
  {ICLR} 2018, Vancouver, BC, Canada, April 30 - May 3, 2018, Workshop Track
  Proceedings}. OpenReview.net, 2018.

\bibitem{9643490}
Sebastian Brandhofer, Ilia Polian, and Hans~Peter Büchler.
\newblock Optimal mapping for near-term quantum architectures based on rydberg
  atoms.
\newblock In {\em 2021 IEEE/ACM International Conference On Computer Aided
  Design (ICCAD)}, pages 1--7, 2021.

\bibitem{Breiman01}
Leo Breiman.
\newblock Random forests.
\newblock {\em Mach. Learn.}, 45(1):5–32, oct 2001.

\bibitem{BreiFrieStonOlsh84}
Leo Breiman, Jerome Friedman, Charles~J. Stone, and R.A. Olshen.
\newblock {\em Classification and Regression Trees}.
\newblock Chapman and Hall/CRC, 1984.

\bibitem{CaiCZYW18}
H.~Cai, T.~Chen, W.~Zhang, Y.~Yu, and J.~Wang.
\newblock Efficient architecture search by network transformation.
\newblock In {\em Proceedings of the Thirty-Second {AAAI} Conference on
  Artificial Intelligence, (AAAI-18), the 30th innovative Applications of
  Artificial Intelligence (IAAI-18), and the 8th {AAAI} Symposium on
  Educational Advances in Artificial Intelligence (EAAI-18), New Orleans,
  Louisiana, USA, February 2-7, 2018}, pages 2787--2794. {AAAI} Press, 2018.

\bibitem{PRXQuantum21}
Lukasz Cincio, Kenneth Rudinger, Mohan Sarovar, and Patrick~J. Coles.
\newblock Machine learning of noise-resilient quantum circuits.
\newblock {\em PRXQuantum21}, 2:010324, Feb 2021.

\bibitem{Datta22}
Kamalika Datta, Abhoy Kole, Indranil Sengupta, and Rolf Drechsler.
\newblock Mapping quantum circuits to 2-dimensional quantum architectures.
\newblock INFORMATIK 2022, 2022.

\bibitem{Dat22}
Kamalika Datta, Abhoy Kole, Indranil Sengupta, and Rolf Drechsler.
\newblock Nearest neighbor mapping of quantum circuits to two-dimensional
  hexagonal qubit architecture.
\newblock In {\em 2022 IEEE 52nd International Symposium on Multiple-Valued
  Logic (ISMVL)}, pages 35--42, 2022.

\bibitem{deMoura08}
Leonardo de~Moura and Nikolaj Bj{\o}rner.
\newblock Z3: An efficient smt solver.
\newblock In C.~R. Ramakrishnan and Jakob Rehof, editors, {\em Tools and
  Algorithms for the Construction and Analysis of Systems}, pages 337--340,
  Berlin, Heidelberg, 2008. Springer Berlin Heidelberg.

\bibitem{9870269}
Lukas Franken, Bogdan Georgiev, Sascha Mucke, Moritz Wolter, Raoul Heese,
  Christian Bauckhage, and Nico Piatkowski.
\newblock Quantum circuit evolution on nisq devices.
\newblock In {\em 2022 IEEE Congress on Evolutionary Computation (CEC)}, pages
  1--8, 2022.

\bibitem{george1993variable}
E.~George and R.~E. McCulloch.
\newblock Variable selection via {Gibbs} sampling.
\newblock {\em Journal of the American Statistical Association},
  88(423):881--889, 1993.

\bibitem{He_Deng_Zheng_Li_Situ_2024}
Zhimin He, Maijie Deng, Shenggen Zheng, Lvzhou Li, and Haozhen Situ.
\newblock Training-free quantum architecture search.
\newblock {\em Proceedings of the AAAI Conference on Artificial Intelligence},
  38(11):12430--12438, Mar. 2024.

\bibitem{Hopcroft79}
John~E. Hopcroft and Jeff~D. Ullman.
\newblock {\em Introduction to Automata Theory, Languages, and Computation}.
\newblock Addison-Wesley Publishing Company, 1979.

\bibitem{qiskit2024}
Ali Javadi-Abhari, Matthew Treinish, Kevin Krsulich, Christopher~J. Wood, Jake
  Lishman, Julien Gacon, Simon Martiel, Paul~D. Nation, Lev~S. Bishop,
  Andrew~W. Cross, Blake~R. Johnson, and Jay~M. Gambetta.
\newblock Quantum computing with {Q}iskit, 2024.

\bibitem{10.5555/1206629}
Phillip Kaye, Raymond Laflamme, and Michele Mosca.
\newblock {\em An Introduction to Quantum Computing}.
\newblock Oxford University Press, Inc., USA, 2007.

\bibitem{Kremer24IBM}
David Kremer, Victor Villar, Hanhee Paik, Ivan Duran, Ismael Faro, and Juan
  Cruz-Benito.
\newblock Practical and efficient quantum circuit synthesis and transpiling
  with reinforcement learning, 2024.

\bibitem{PhysRevLett116.230504}
U.~Las~Heras, U.~Alvarez-Rodriguez, E.~Solano, and M.~Sanz.
\newblock Genetic algorithms for digital quantum simulations.
\newblock {\em Phys. Rev. Lett.}, 116:230504, Jun 2016.

\bibitem{PhysRevResearchLi20}
Li~Li, Minjie Fan, Marc Coram, Patrick Riley, and Stefan Leichenauer.
\newblock Quantum optimization with a novel gibbs objective function and ansatz
  architecture search.
\newblock {\em Phys. Rev. Res.}, 2:023074, Apr 2020.

\bibitem{Li24}
Yangzhi Li, Wen Liu, and Maoduo Li.
\newblock Deep reinforcement learning for mapping quantum circuits to 2d
  nearest-neighbor architectures.
\newblock {\em Advanced Quantum Technologies}, 7(2):2300289, 2024.

\bibitem{Marcus1967}
Mitchell~P. Marcus.
\newblock {\em Switching Circuits for Engineers}.
\newblock Prentice-Hall, 1967.

\bibitem{martyniuk2024quantumarchitecturesearchsurvey}
Darya Martyniuk, Johannes Jung, and Adrian Paschke.
\newblock Quantum architecture search: A survey, 2024.

\bibitem{doi:10.1080/01621459.1949.10483310}
Nicholas Metropolis and S.~Ulam.
\newblock The monte carlo method.
\newblock {\em Journal of the American Statistical Association},
  44(247):335--341, 1949.

\bibitem{Miikkulainen2020}
Risto Miikkulainen.
\newblock {\em Neuroevolution}, pages 1--8.
\newblock Springer US, New York, NY, 2020.

\bibitem{9138945}
Prakash Murali, Dripto~M. Debroy, Kenneth~R. Brown, and Margaret Martonosi.
\newblock Architecting noisy intermediate-scale trapped ion quantum computers.
\newblock In {\em 2020 ACM/IEEE 47th Annual International Symposium on Computer
  Architecture (ISCA)}, pages 529--542, 2020.

\bibitem{negrinho2019towards}
R.~Negrinho, D.~Patil, N.~Le, D.~Ferreira, M.~Gormley, and G.~Gordon.
\newblock Towards modular and programmable architecture search.
\newblock {\em Neural Information Processing Systems}, 2019.

\bibitem{nielsen2010quantum}
Michael~A Nielsen and Isaac~L Chuang.
\newblock {\em Quantum computation and quantum information}.
\newblock Cambridge university press, 2010.

\bibitem{Note1}
\protect \url
  {https://github.com/Qiskit/qiskit/blob/main/qiskit/transpiler/preset_passmanagers/level3.py}
  (accessed 10/2024).

\bibitem{Note2}
\protect \url {https://quantum.ibm.com} (accessed 08/2024).

\bibitem{Note3}
\protect \url {https://quantum.ibm.com/services/resources?type=Eagle} (accessed
  08/2024).

\bibitem{Ovide23}
Anabel Ovide, Rodrigo Santiago, Medina Bandic, Hans van Someren, Sebastian
  Feld, Sergi Abadal, Eduard Alarcon, and Carmen Almudever.
\newblock Mapping quantum algorithms to multi-core quantum computing
  architectures.
\newblock In {\em IEEE International Symposium on Circuits and Systems
  (ISCAS)}, pages 1--5, 05 2023.

\bibitem{Quinlan1986InductionOD}
J.~Ross Quinlan.
\newblock Induction of decision trees.
\newblock {\em Machine Learning}, 1:81--106, 1986.

\bibitem{RasconiOddi2019}
Riccardo Rasconi and Angelo Oddi.
\newblock An innovative genetic algorithm for the quantum circuit compilation
  problem.
\newblock {\em Proceedings of the AAAI Conference on Artificial Intelligence},
  33:7707--7714, Jul. 2019.

\bibitem{riu2024RLZX}
Jordi Riu, Jan Nogué, Gerard Vilaplana, Artur Garcia-Saez, and Marta~P.
  Estarellas.
\newblock Reinforcement learning based quantum circuit optimization via
  zx-calculus, 2024.

\bibitem{PhysRevA.110.022443}
Bodo Rosenhahn and Christoph Hirche.
\newblock Quantum normalizing flows for anomaly detection.
\newblock {\em Phys. Rev. A}, 110:022443, Aug 2024.

\bibitem{RosOsb2023a}
Bodo Rosenhahn and Tobias~J. Osborne.
\newblock Monte carlo graph search for quantum circuit optimization.
\newblock {\em Phys. Rev. A}, 108:062615, Dec 2023.

\bibitem{Ryabov15}
V.~A. Ryabov.
\newblock Quantum volume.
\newblock {\em International Journal of Modern Physics B}, 29(23):1550166,
  2015.

\bibitem{Stefano2024}
Manuel~De Stefano, Dario~Di Nucci, Fabio Palomba, and Andrea~De Lucia.
\newblock An empirical study into the effects of transpilation on quantum
  circuit smells.
\newblock {\em Empirical Software Engineering}, 29(3):61, May 2024.

\bibitem{Top23}
Rasit~O. Topaloglu, editor.
\newblock {\em Design Automation of Quantum Computers}.
\newblock Springer Nature, Switzerland AG 2023, 2023.

\bibitem{vandewetering2020zx}
John van~de Wetering.
\newblock Zx-calculus for the working quantum computer scientist, 2020.

\bibitem{Leuuwen91}
Jan van Leeuwen, editor.
\newblock {\em Handbook of theoretical computer science (vol. B): formal models
  and semantics}.
\newblock MIT Press, Cambridge, MA, USA, 1991.

\bibitem{WangAQC23}
Peiyong Wang, Muhammad Usman, Udaya Parampalli, Lloyd C.~L. Hollenberg, and
  Casey~R. Myers.
\newblock Automated quantum circuit design with nested monte carlo tree search.
\newblock {\em IEEE Transactions on Quantum Engineering}, pages 1--25, 2023.

\bibitem{quantum3020021}
Masaya Watabe, Kodai Shiba, Chih-Chieh Chen, Masaru Sogabe, Katsuyoshi
  Sakamoto, and Tomah Sogabe.
\newblock Quantum circuit learning with error backpropagation algorithm and
  experimental implementation.
\newblock {\em Quantum Reports}, 3(2):333--349, 2021.

\bibitem{PhysRevResearch.2.033446}
Matteo~M. Wauters, Emanuele Panizon, Glen~B. Mbeng, and Giuseppe~E. Santoro.
\newblock Reinforcement-learning-assisted quantum optimization.
\newblock {\em Phys. Rev. Res.}, 2:033446, Sep 2020.

\bibitem{xie2018snas}
Sirui Xie, Hehui Zheng, Chunxiao Liu, and Liang Lin.
\newblock {SNAS}: stochastic neural architecture search.
\newblock In {\em International Conference on Learning Representations}, 2019.

\bibitem{Zhang2021}
Shi-Xin Zhang, Chang-Yu Hsieh, Shengyu Zhang, and Hong Yao.
\newblock Neural predictor based quantum architecture search.
\newblock {\em Machine Learning: Science and Technology}, 2(4):045027, oct
  2021.

\bibitem{Zhang2022}
Shi-Xin Zhang, Chang-Yu Hsieh, Shengyu Zhang, and Hong Yao.
\newblock Differentiable quantum architecture search.
\newblock {\em Quantum Science and Technology}, 7(4):045023, aug 2022.

\bibitem{PhysRevX10021067}
Leo Zhou, Sheng-Tao Wang, Soonwon Choi, Hannes Pichler, and Mikhail~D. Lukin.
\newblock Quantum approximate optimization algorithm: Performance, mechanism,
  and implementation on near-term devices.
\newblock {\em Phys. Rev. X}, 10:021067, Jun 2020.

\bibitem{Zhu23ICACS}
Weiwei Zhu, Jiangtao Pi, and Qiuyuan Peng.
\newblock A brief survey of quantum architecture search.
\newblock In {\em Proceedings of the 6th International Conference on
  Algorithms, Computing and Systems}, ICACS '22, New York, NY, USA, 2023.
  Association for Computing Machinery.

\bibitem{zulehner2018mapping}
Alwin Zulehner, Alexandru Paler, and Robert Wille.
\newblock An efficient methodology for mapping quantum circuits to the ibm qx
  architectures.
\newblock {\em IEEE Transactions on Computer Aided Design of Integrated
  Circuits and Systems (TCAD)}, 2018.

\end{thebibliography}

\end{document}